\documentclass[aps,prl,reprint,showpacs,superscriptaddress,longbibliography,nourl, noeprint]{revtex4-2}
\usepackage{amsmath,amssymb,physics,bm}
\usepackage{xcolor,soul,setspace}
\usepackage{graphicx,microtype,siunitx} 
\usepackage[colorlinks=true,linkcolor=blue,citecolor=blue,urlcolor=blue]{hyperref} 
\usepackage[all]{hypcap} 

\newcommand{\kB}[0]{k_{\rm B}}
\newcommand{\rvec}[0]{{\bm r}} 
\newcommand{\nvec}[0]{{\bm n}}
\newcommand{\ua}[0]{v_{\rm a}} 
\newcommand{\Dr}[0]{D_\mathrm{r}}
\newcommand{\Dc}[0]{D_{\rm c}} 
\newcommand{\muc}[0]{\mu_{\rm c}}
\newcommand{\RH}[0]{R_{\rm h}}
\newcommand{\fx}[0]{f_\parallel}
\newcommand{\fy}[0]{f_\perp}
\newcommand{\Deff}[0]{\tilde{D}} 
\newcommand{\mueff}[0]{\tilde{\mu}}
\newcommand{\xic}[0]{\xi_{\rm c}}
\newcommand{\xir}[0]{\xi_{\rm r}} 
\newcommand{\Dctr}[0]{\bar{D}_{\rm c}} 
\newcommand{\Dtr}[0]{\bar{D}} 
\newcommand{\Votr}[0]{\bar{V}_0} 
\newcommand{\Ftr}[0]{\bar{F}} 
\newcommand{\fxtr}[0]{\bar{f}_\parallel} 
\newcommand{\fytr}[0]{\bar{f}_\perp} 
\newcommand{\uatr}[0]{\bar{v}_{\rm a}} 
\newcommand{\utr}[0]{\bar{u}} 
\newcommand{\zetac}[0]{\zeta_{\rm c}}
\newcommand{\zetar}[0]{\zeta_{\rm r}} 
\newcommand{\Nc}[0]{\mathcal{N}_{\rm c}(0,1)}
\newcommand{\Nr}[0]{\mathcal{N}_{\rm r}(0,1)} 
\newcommand{\Nx}[0]{\mathcal{N}_x(0,1)} 
\newcommand{\Ny}[0]{\mathcal{N}_y(0,1)} 
\begin{document} 
\title[]{Mechanochemical Active Ratchet}

\author{Artem Ryabov}
\email[]{rjabov.a@gmail.com}
\affiliation{Charles University, Faculty of Mathematics and Physics, Department of Macromolecular Physics, V Hole\v{s}ovi\v{c}k\'ach 2, CZ-18000 Praha 8, Czech Republic}
\affiliation{Departamento de F\'{\i}sica, Faculdade de Ci\^{e}ncias, Universidade de Lisboa, 1749-016 Lisboa, Portugal}
\affiliation{Centro de F\'{i}sica Te\'{o}rica e Computacional, Faculdade de Ci\^{e}ncias, Universidade de Lisboa, 1749-016 Lisboa, Portugal}
\author{Mykola Tasinkevych}
\email[]{mykola.tasinkevych@ntu.ac.uk}
\affiliation{Departamento de F\'{\i}sica, Faculdade de Ci\^{e}ncias, Universidade de Lisboa, 1749-016 Lisboa, Portugal}
\affiliation{Centro de F\'{i}sica Te\'{o}rica e Computacional, Faculdade de Ci\^{e}ncias, Universidade de Lisboa, 1749-016 Lisboa, Portugal} 
\affiliation{SOFT Group, School of Science and Technology, Nottingham Trent University, Clifton Lane, Nottingham NG11~8NS, United Kingdom}
\affiliation{International Institute for Sustainability with Knotted Chiral Meta Matter, Hiroshima University, Higashihiroshima 739-8511, Japan.}

\date{November 10, 2023}

\begin{abstract}
\noindent{\bf \large Abstract} 

\noindent Self-propelled nanoparticles moving through liquids offer the possibility of creating advanced applications where such nanoswimmers can operate as artificial molecular-sized motors. Achieving control over the motion of nanoswimmers is a crucial aspect for their reliable functioning. While the directionality of micron-sized swimmers can be controlled with great precision, steering nano-sized active particles poses a real challenge. One of the reasons is the existence of large fluctuations of active velocity at the nanoscale. Here, we describe a mechanism that, in the presence of a ratchet potential, transforms these fluctuations into a net current of active nanoparticles. We demonstrate the effect using a generic model of self-propulsion powered by chemical reactions. The net motion along the easy direction of the ratchet potential arises from the coupling of chemical and mechanical processes and is triggered by a constant, transverse to the ratchet, force. The current magnitude sensitively depends on the amplitude and the periodicity of the ratchet potential and the strength of the transverse force. Our results highlight the importance of thermodynamically consistent modeling of chemical reactions in active matter at the nanoscale and suggest new ways of controlling dynamics in such systems. 
\end{abstract}
\maketitle  

\noindent {\large \bf Introduction}

\noindent
Chemistry and physics of micron- and nano-sized self-propelled particles are rapidly growing fields exploring optimal designs and fundamental concepts of self-propulsion mechanisms. A current intense discussion focuses on a possible self-propulsion of individual catalytic macromolecules that could serve as basic components for artificial nanomachines \cite{Patino/etal:2018,Zhao/etal:2018,Safdar/etal:2018,Huang/etal:2021, Ghosh/etal:2021,Amano/etal:2021, arque/etal:2022, pumm/etal:2022, Shi/etal:2022}. 

The self-propelled (active) motion of micron-sized particles is driven by a large number of chemical reactions and/or hydrodynamic flows in an ambient fluid. Relatively large dimensions of these objects lead to negligible fluctuations in the magnitude of their active velocities. This renders the particle dynamics accurately controllable by employing laser fields \cite{Zemanek/etal:2019, Volpe/etal:2023}, concentration gradients \cite{hong/etal:2007,baraban/etal:2013,dey/etal:2013,saha/etal:2014,peng/etal:2015}, local heating \cite{Braun/etal:2015, Baffou/etal:2020}, using external magnetic fields in combination with suitably prepared magnetic particles \cite{kline/etal:2005, solovev/etal:2010, baraban/etal:2012}, or by a specific chemical patterning of surfaces \cite{simmchen/etal:2016,Das/etal:2015,uspal/etal:2016,uspal/etal:2019}. The precise guidance is essential for several applications including targeted drug delivery \cite{li/etal:2016,peng/etal:2015,patra/etal:2013}, biosensing \cite{Wu:2010}, transport of microcargoes in lab-on-chip devices, \cite{baraban/etal:2012b} or assembly of microstructures \cite{sundararajan/etal:2008, sundararajan/etal:2010, sanchez/etal:2015, palagi/etal:2018}. 

In contrast, strong fluctuations are unavoidable at the nanoscale, where dimensions of active particles are on par with those of intracellular molecular motors like kinesins, myosins, and ribosomes \cite{Vale/Milligan:2000, Schliwa/Woehlke:2003, Alberts/etal:2014}, and artificial molecular machines and pumps \cite{Erbas-Cakmak/etal:2015, Feng/etal:2021, Borsley/etal:2022}. Active motion at these scales can be caused by a comparably small number of reactions and is strongly influenced by thermal noise. Large fluctuations of the magnitude and the direction of self-propulsion velocity hinder the development of methods to efficiently control the dynamics of nanoswimmers. 

Here, we use a minimal thermodynamically consistent model of particle's self-propulsion  \cite{Pietzonka/Seifert:2018, Speck:2018, Ryabov/Tasinkevych:SoftMatt2022, Ryabov/Tasinkevych:JCP2022} to demonstrate a novel physical mechanism for rectifying and guiding trajectories of chemical nanoswimmers. The mechanism benefits from the presence of strong fluctuations in the direction of self-propulsion velocity and it arises as a direct consequence of the main premise that the chemical reactions driving the active motion comply with the principle of microscopic reversibility (MR) \cite{Tolman:PNAS1925, Onsager:1931a, Blackmond:2009, Astumian:2016}. The resulting rectification effect can be used to guide and sort active nanoparticles based on their propulsion mechanisms, because a corresponding reference model without the MR, as well as passive Brownian particles, does not exhibit this type of rectified motion. 
\vspace{0.5cm}

\begin{figure}[b]
\centering
\includegraphics[width=\columnwidth]{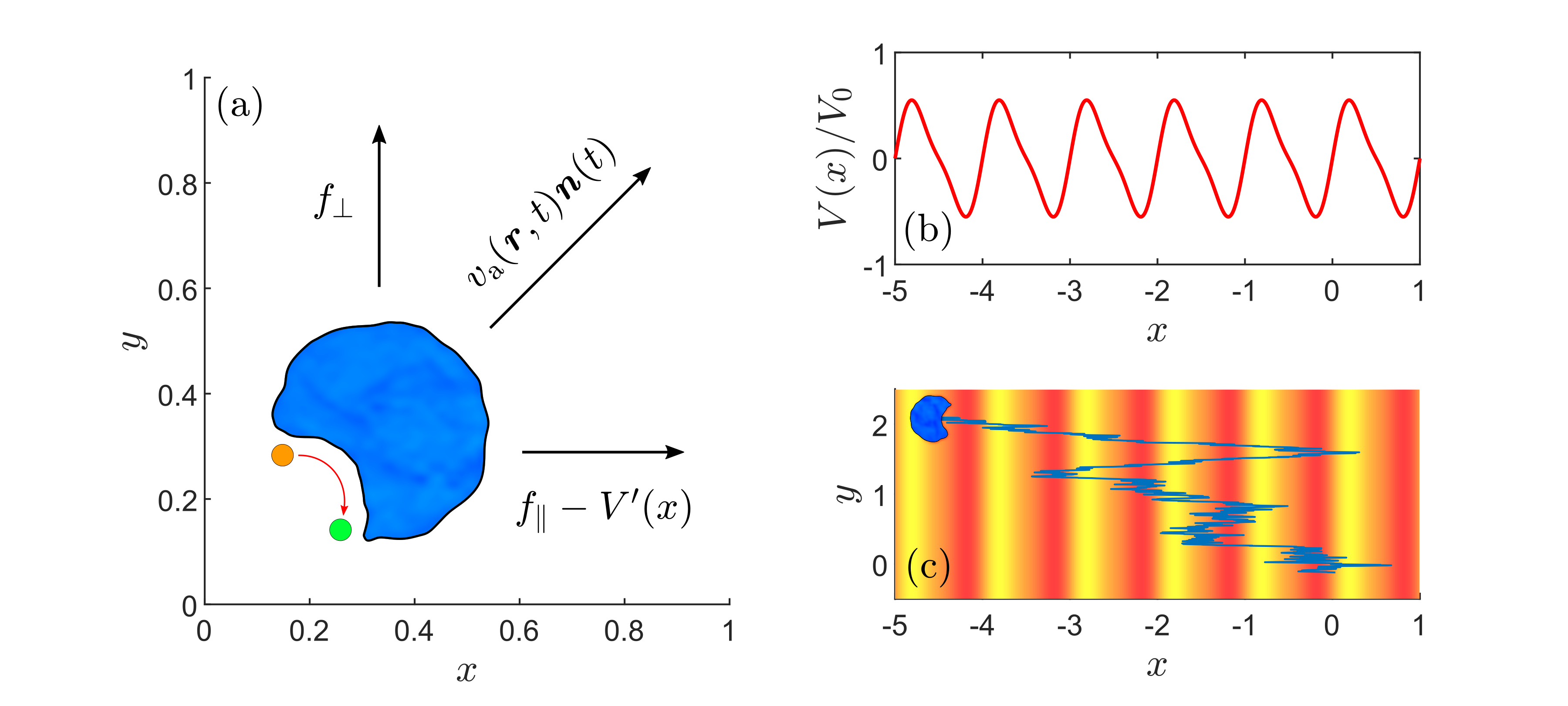} 
\caption{{\bf Illustration of the model.} (a) An active nanoparticle (blue) whose self-propelled motion with speed $\ua(\rvec,t)$ occurs in the direction of orientation $\nvec(t)$ (diagonal arrow) and is powered by chemical reactions (red arrow). Assuming the microscopic reversibility of the self-propulsion mechanism gives rise to the mechanochemical coupling term $[\muc \nvec(t) \cdot \bm F(\rvec)]$ in Eq.~\eqref{eq:ua} for $\ua(\rvec ,t)$. The $x$- and $y$-component of the external force $\bm F(\rvec )$, Eq.~\eqref{eq:F}, are represented by the horizontal and vertical arrows. (b) The periodic asymmetric potential $V(x)$ given in Eq.~\eqref{eq:V} normalized by its barrier height $V_0$ for $\lambda=1$. (c) Sketch of a trajectory of the nanoparticle diffusing in the potential $V(x)$ (dark red marks the potential minima, light orange the maxima) and subjected to a constant force with the amplitude $\fy >0$ acting in the $y$-direction and with no constant force component in the $x$-direction ($\fx=0$). The nonzero perpendicular bias $\fy$ can induce a net particle motion along the $x$-axis. This special consequence of the mechanochemical coupling term in $\ua(\rvec,t)$ is illustrated in Fig.~\ref{fig:vx_fx-0}. If $\fx >0$, the direction of the motion can be reversed by changing $\fy$ as shown in Fig.~\ref{fig:vx-reversal}.}
\label{fig:model}
\end{figure} 

\noindent {\large \bf Results and Discussion}

\noindent Stochastic Markovian models become thermodynamically consistent if rates of elementary transitions obey the local detailed balance~\cite{Maes:SciPost2021}. For an overdamped Brownian motion subjected to an external force ${\bm F}(\rvec)$ and to thermal fluctuations represented by the zero-mean Gaussian white noise ${\bm \xi}(t)$, the local detailed balance condition is included in the Langevin equation for the particle velocity, 
$\dot \rvec =  \mu {\bm F}(\rvec) + \sqrt{2 D }\, {\bm \xi}(t)$, by requiring that the translational diffusion constant $D$ is related to the mobility $\mu$ via the fluctuation-dissipation relation $D= \mu \kB T$, $T$ being the temperature of ambient fluid. 

In addition to the force- and thermally-driven movements, the velocity of a self-propelled nanoparticle is influenced by its chemically-powered active dynamics. The particle center of mass position $\rvec(t) = (x(t),y(t))$ then obeys the Langevin equation 
\begin{equation}
\label{eq:Langevin-general}
\frac{\dd \rvec }{\dd  t} = \ua (\rvec,t) \nvec(t) + \mu {\bm F}(\rvec) + \sqrt{2 D }\, {\bm \xi}(t), 
\end{equation} 
where the first term on the right-hand side represents the active velocity of magnitude $\ua (\rvec,t)$ oriented in the direction of $\nvec(t)$, see Fig.~\ref{fig:model}(a). 

The vector ${\bm n}(t)=( \cos\phi(t) ,\sin\phi(t) )$, giving the orientation of the nanoparticle, performs  rotational diffusion \cite{Han/etal:2006}, i.e., the angle $\phi(t)$ undergoes the Brownian motion 
\begin{equation} 
\label{eq:Langevin-rotational}
\dot \phi(t) = \sqrt{2 \Dr }\, \xir(t) , 
\end{equation} 
 with $\Dr$ being the rotational diffusion constant and $\xir(t)$ a zero-mean Gaussian white noise. 

The speed $\ua (\rvec,t)$ corresponds to the chemically-driven particle motion in the direction of ${\bm n}(t)$. If the kinetics of the underlying reactions comply with the principle of MR, then, on the mesoscopic timescale, where an infinitesimal time interval $\dd t$ covers many chemically-driven particle jumps, we may approximate the speed by \cite{Ryabov/Tasinkevych:SoftMatt2022} 
\begin{equation}
\label{eq:ua}
\ua(\rvec,t) \approx u + \muc \nvec (t) \cdot {\bm F}(\rvec) + \sqrt{2 \Dc}\, \xic(t). 
\end{equation} 
Here, $u$ is the constant mean active speed in the zero-force case (${\bm F}=0$), the dot $\cdot$ denotes the scalar product, and $\xic(t)$ is a Gaussian white noise stemming from fluctuations of the number of chemical reactions per unit time. Positive constants $\Dc$ and $\muc$ are related by $\Dc = \muc \kB T $. 
The white noises $\xi_\alpha(t)$ in Eqs.~\eqref{eq:Langevin-general}-\eqref{eq:ua} satisfy 
$\langle \xi_\alpha(t) \rangle =0$, and 
$\langle \xi_\alpha(t) \xi_\beta(t') \rangle = \delta_{\alpha\beta} \delta(t-t')$,
$\alpha, \beta \in \{x,y,\textrm{r},\textrm{c}\}$.
A derivation of Eq.~\eqref{eq:ua} starting from the mesoscopic jump-diffusion model can be found in Ref.~\cite{Ryabov/Tasinkevych:SoftMatt2022}. The central assumption of this derivation is that the rates of motion-powering chemical reactions satisfy the local detailed balance condition. This is equivalent to assuming MR of the kinetics of these reactions, which means that their underlying microscopic dynamics are symmetric with respect to the time-reversal \cite{Maes:SciPost2021}. A general aim of the current work is to demonstrate the impact of individual terms in~\eqref{eq:ua} on the dynamics of an active nanoparticle diffusing in external force fields.

According to Eq.~\eqref{eq:ua}, the active speed depends on the component $\nvec (t) \cdot {\bm F}(\rvec)$ of the external force along $\nvec (t)$. This term describes the mechanochemical coupling \cite{Beyer/Clausen-Schaumann:2005} and allows to sculpt a free energy landscape of the macromolecule by the action of external forces. The coupling is utilized in various single-molecule force spectroscopies \cite{Bustamante/etal:2003, Bustamante/etal:2021, Volpe/etal:2023}, biosensing \cite{Hu/etal:2022} and underlies the ability of molecular motors to exert forces and torques \cite{Astumian/Bier:1996, Toyabe/etal:2011, Mukherjee/etal:2015, Trivedi/etal:2020, Feng/etal:2021, Borsley/etal:2021}. 

The force-dependence of $\ua(\rvec,t)$ is a direct consequence of MR of the self-propulsion mechanism and it has prominent impacts on particle dynamics \cite{Ryabov/Tasinkevych:SoftMatt2022, Ryabov/Tasinkevych:JCP2022}. We shall use this dependence to design a new method of nanoparticle guiding, which is inaccessible in systems lacking the mechanochemical coupling, like in the case of micron-sized particles with constant active speed, see Eq.~\eqref{eq:uABP} below. In particular, we report on a ratchet effect that occurs if the particle is subjected to the external force 
\begin{equation}
\label{eq:F}
\bm F(\rvec) = \left(\fx -V'(x), \fy \right) ,
\end{equation} 
determined by the constant components $\fx$ and $\fy$, and by the derivative of the asymmetric $\lambda$-periodic potential 
\begin{equation} 
\label{eq:V}
V(x) = \frac{V_0}{2} \left[\sin\left(\frac{2\pi x}{\lambda} \right) + \frac{1}{4} \sin\left(\frac{4\pi x}{\lambda}\right) \right] , 
\end{equation}   
shown in Fig.~\ref{fig:model}(b). A trajectory of the particle moving in $V(x)$ and being acted upon by $\fy >0$ is sketched in Fig.~\ref{fig:model}(c). 

The asymmetric sawtooth-like potential $V(x)$ is qualitatively similar to potentials used in various theoretical and experimental studies of ratchets \cite{Reimann:2002, Angelani/etal:2011, Ai/Wu:2014, McDermott/etal:2016, Ryabov/etal:2016, Arzola/etal:2011, Lozano/etal:2016, Arzola/etal:2017, Skaug/etal:2018, Schwemmer/eta:2018, Stoop/etal:2019, Paneru/etal:2021, Leyva/etal:2022}. A distinctive feature of the ratchet mechanism discussed below is that the emergent current of particles in the $x$-direction is induced by the transverse force $\fy$ applied perpendicular to it in the $y$-direction. Although here we focus on the two-dimensional case, in an actual experiment, the force $\fy$ can point in any direction in the $yz$-plane.

Assuming that the force in Eq.~\eqref{eq:F} can be accurately adjusted in an experiment, we discuss the dependence of the ratchet effect on $\fx$, $\fy$, and $V_0$. The phenomenological model parameters $D$, $\Dr$, $\Dc$, and $u$  will be kept fixed. All results reported below are obtained assuming $\lambda = 100\;\si{nm}$. To get the results, we have integrated Eqs.~\eqref{eq:Langevin-general} and~\eqref{eq:Langevin-rotational} numerically, see Supplementary Note~1 and~2 for technical details. 

The values of the diffusion constants $D$ and $\Dr$ were chosen having catalytically active enzymes in mind \cite{Jee/etal:PNAS2018, Ah-Young/etal:2018, Jee:2020}. Thus, $D$ was approximated by the Stokes-Einstein equation $D = \kB T/6\pi\eta \RH$, and $\Dr = \kB T/8\pi\eta \RH^3$ valid for a sphere of radius $\RH$, $\eta$ is the fluid dynamic viscosity. With $\RH=15\;\si{nm}$,  $T = 300\;\si{K}$, 
and $\eta = 8.53\times  10^{-4}\;\si{Ns/m^2}$, which corresponds to the dynamic viscosity of water at this temperature, we find  
\begin{align} 
\label{eq:D}
D & \approx 1.7\times 10^{-11}\;\si{m^2/s},\\  
\Dr & \approx 5.7\times 10^{4}\;\si{s^{-1}}.     
\label{eq:Dr}
\end{align}

Contrary to $D$ and $\Dr$, the behavior of the phenomenological parameter $\Dc$ remains poorly understood. To approximate its value, we have used $\Dc \approx (\delta r)^2 (k_++k_-)/2$ \cite{Ryabov/Tasinkevych:SoftMatt2022}, where $k_+$ is the rate of the reaction causing the nanoparticle displacement $\delta r {\bm n}$ (at $\bm F=0$), and neglected the rate $k_-$ of the reversed reaction, which causes the displacement by $-\delta r {\bm n}$. For $k_+ = 10^5\;\si{s^{-1}}$ and $\delta r = 5\;\si{nm}$,  
\begin{equation} 
\label{eq:Dc}
\Dc\approx 1.3\times 10^{-12}\;\si{m^2/s}.  
\end{equation} 
We note that the very existence of the reported ratchet effect follows from the condition $\Dc >0$ and does not depend on the actual value of $\Dc$. Equation~\eqref{eq:Langevin-x-approx} derived below will demonstrate that the ratios $\Dc \fy /\kB T$ and $\mu V_0/\lambda$ are the two key parameters that control the occurrence and the strength of the ratchet effect. Their values can be changed in an experiment by adjusting the force $\fy$, the barrier height $V_0$, and the period $\lambda$. 

For the constant part of the speed~\eqref{eq:ua}, we have used $u=10\;\si{nm/s}$. The reported results, however, are insensitive to the actual value of $u$, as we have checked several alternative values of $u$  between 0 and $10^3\;\si{nm/s}$. The negligible effect of $u$ on the translational dynamics results from the large $\Dr$, Eq.~\eqref{eq:Dr}, characteristic for small particles ($\Dr \sim 1/\RH^3$) as compared to the slower translational motion whose speed is determined by $D$ in~\eqref{eq:D} ($D \sim 1/\RH$),  $u$ and by $\Dc$~\eqref{eq:Dc}. This is in sharp contrast to the motion of micron-sized active particles that move ballistically for significant periods of time due to relatively large $u$ and small $\Dr$ \cite{Howse/etal:2007, tenHagen/etal:2011, Dunderdale/etal:2012, Romanczuk2012, Zottl/Stark:JPCM2016, Choudhury/etal:2017, Patino/etal:2018}.

We quantify the ratchet effect in terms of the mean velocity $\langle v_x \rangle$ of the particle motion in the $x$-direction. It is defined as the average 
\begin{equation}
\langle v_x \rangle = \frac{1}{N_{\rm tr}}\sum_{i=1}^{N_{\rm tr}} \frac{x_i(t_{\rm max})}{t_{\rm max}}
\end{equation} 
over $N_{\rm tr}$ simulated trajectories, where $x_i(t_{\rm max})$ is the final $x$-coordinate of the $i$-th trajectory. We have used $\Dr t_{\rm max}=10^3$, $N_{\rm tr}=10^4$, and checked that increasing $t_{\rm max}$ and  $N_{\rm tr}$ does not affect the displayed results.

\begin{figure}[t!]
\centering
\includegraphics[width=0.9\columnwidth]{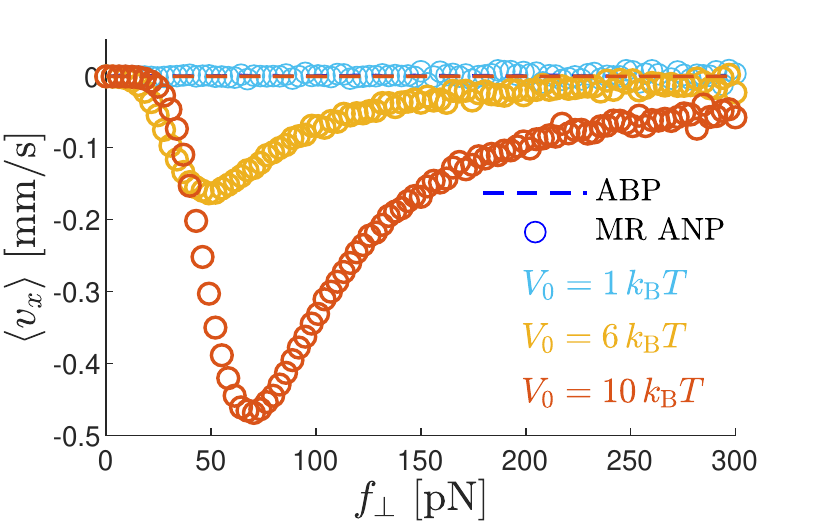} 
\caption{{\bf Rectified particle motion.} Net particle translation induced by the mechanochemical coupling term in the active speed~\eqref{eq:ua}. The mean particle velocity $\langle v_x \rangle$ in the $x$-direction is plotted as the function of the perpendicular force $\fy$ acting in the $y$-direction for three values of the barrier height $V_0$ (colors) and $\fx=0$. Parameters $D$, $\Dr$, $\Dc$ are given in Eqs.~\eqref{eq:D}-\eqref{eq:Dc}, $\lambda = 100\;\si{nm}$, and $u=10\;\si{nm/s}$. Symbols represent simulated $\langle v_x \rangle$ of the active nanoparticle with the microscopically reversible self-propulsion mechanism (MR~ANP) illustrated in Fig.~\ref{fig:model}. Dashed lines show corresponding results for the active Brownian particle (ABP) model with constant active speed~\eqref{eq:uABP}. The three  dashed lines overlap.}
\label{fig:vx_fx-0}
\end{figure} 

Figure~\ref{fig:vx_fx-0} shows $\langle v_x \rangle$ for $\fx=0$, i.e., when there is no net force in the $x$-direction. Symbols mark $\langle v_x \rangle$ of the active nanoparticle with the microscopically reversible self-propulsion mechanism (MR~ANP) described by the active speed $\ua(\rvec ,t)$ in~\eqref{eq:ua}. Dashed lines represent $\langle v_x \rangle$ of the so-called active Brownian particle (ABP) model \cite{Erdmann/etal:EPJB2000, Szabo/etal:PRE2006, Peruani/Morelli:PRL2007, Teeffelen/Loewen:PRE2008, tenHagen/etal:2011, Henkes/etal:2011, Bialke/etal:PRL2012, Romanczuk2012, Pototsky/Stark:EPL2012,  Buttinoni/etal:2013, Yang/etal:SOFTMATTER2014, Stenhammar/etal:SOFTMATTER2014, Zottl/Stark:JPCM2016,  Das/etal:NJP2018,  Malakar/etal:2020, Chaudhuri/Dhar:JSTAT2021} 
which is a paradigmatic model for micron-sized active colloidal particles \cite{Howse/etal:2007, tenHagen/etal:2011, Dunderdale/etal:2012, Romanczuk2012, Zottl/Stark:JPCM2016, Patino/etal:2018}  and is characterized by the constant active speed  
\begin{equation}
\label{eq:uABP} 
\ua^{({\rm ABP})} = u.
\end{equation} 
The ABP model is formally obtained from the MR~ANP model after neglecting the mechanochemical coupling and the noise in Eq.~\eqref{eq:ua}, i.e., by setting $\muc=\Dc/\kB T=0$. 

The simulated mean velocities $\langle v_x \rangle$  for these two models are plotted in Fig.~\ref{fig:vx_fx-0} as the functions of the perpendicular force $\fy$ for three representative values of the barrier height~$V_0$:  

(i) $V_0 = 1\; \kB T$,  light-blue symbols, where the barrier height is equal to the energy of thermal fluctuations. Thus, the asymmetric potential $V(x)$ has only a minor influence on the dynamics and $\langle v_x \rangle$ of MR~ANP is vanishingly small.  

(ii) $V_0 = 6\;\kB T$, yellow symbols. Here, thermal fluctuations can induce hopping-like transitions of particles over the barriers of $V(x)$. As a result of the ratchet effect, MR~ANP exhibits a nonzero mean velocity in the negative direction of the $x$-axis ($\langle v_x \rangle \leq 0$). There exists an optimal value of $\fy$ leading to a maximal negative velocity of MR~ANP. In the limits of small ($\fy\to 0$) and large ($\fy\to \infty$) perpendicular forces, $\langle v_x \rangle$ vanishes. 

(iii) $V_0 = 10\; \kB T$, red symbols. In this case, the barriers of~$V(x)$ are large compared to the energy of thermal fluctuations. The cross-well transitions of MR~ANP are driven dominantly by the active velocity. The ratchet effect becomes the most pronounced out of the three cases. 

In Fig.~\ref{fig:vx_fx-0}, the three dashed lines marking $\langle v_x \rangle$ for the ABP model overlap and follow closely the line $\langle v_x \rangle =0$ for all $\fy$. 

These observations suggest that the ratchet effect should be caused by the force-dependence of the active speed~\eqref{eq:ua}. 
Can we understand this effect qualitatively, e.g., by relating it to some of the well-understood ratchet mechanisms? It turns out that this is not only possible but is indeed an insightful step as it will lead to yet another way of control over the motion of the nanoparticle. 

For the considered values of parameters, the Langevin equation for the $x$-coordinate of MR~ANP (and for $\fx=0$)  approximately reduces to 
\begin{equation} 
\label{eq:Langevin-x-approx}
\frac{\dd x}{\dd t} \approx  
\frac{\muc \fy}{2}\sin[2\phi(t)] 
- \mueff \frac{\dd V}{\dd x} 
+ \sqrt{2\Deff}\, \xi_x(t),  
\end{equation} 
where the effective enhanced mobility $\mueff = \mu +\muc/2$, and the enhanced diffusivity $\Deff = D+\Dc/2$ (compared to $\mu$ and $D$), satisfy the fluctuation-dissipation relation $\Deff = \mueff \kB T$. 
Indeed, due to the fast rotational diffusion, we can neglect $u$ in Eq.~\eqref{eq:ua} as discussed above, yielding  
$\ua(\rvec ,t)\cos\phi \approx \muc \fy \sin (2\phi)/2 - \muc V'(x) \cos^2\!\phi + \cos\phi\sqrt{2\Dc}\,\xic $, for the $x$-component of the active velocity. Since $\phi$ is a fast process and $\mu \gg \muc$, $\cos^2\!\phi$ can be replaced by its mean value,  $\cos^2\phi \approx 1/2$. The noise term is then simplified accordingly: $ \cos\phi\,\sqrt{2\Dc}\,\xic = \sqrt{2\Dc\cos^2\!\phi}\, \xic \approx \sqrt{\Dc}\, \xic$, and is added to the thermal noise $\sqrt{2D}\, \xi_x$, which eventually leads to~\eqref{eq:Langevin-x-approx};  see Supplementary Note~3 for more technical details and tests of validity of this approximation.

Equation~\eqref{eq:Langevin-x-approx} describes the dynamics of a basic model of a Brownian motor, the so-called overdamped tilting ratchet \cite{Astumian:1997, Reimann:2002, Arzola/etal:2011, Arzola/etal:2017, Reichhardt/etal:2017}. Discussion of the dynamics governed by Eq.~\eqref{eq:Langevin-x-approx} can provide a further physical understanding of the mechanism leading to the reported ratchet effect. Below, we focus on qualitative arguments and intuitive interpretation of the effect. For a thorough mathematical analysis of various classes of tilting ratchet models, we refer to Sec.~5 of the review article~\cite{Reimann:2002}.

In particular, Eq.~\eqref{eq:Langevin-x-approx} describes the dynamics of an overdamped Brownian particle diffusing in the potential $V(x)$ subject to the time-dependent stochastic force $(\fy/2)\sin[2\phi(t)]$. If the magnitude of the force becomes large enough, the particle may start surmounting potential barriers of $V(x)$ located in the direction of the force. Because $V(x)$ is asymmetric, shapes of the left (moderate slope) and right (steeper slope) barrier surrounding a given well differ and the barrier-crossing is easier in one of the two directions: in $(-x)$-direction in the present case. After surpassing multiple barriers in both directions, the asymmetry of $V(x)$  translates into the nonzero mean motion in the easy $(-x)$-direction giving a nonzero mean velocity $\langle v_x \rangle$ observed in Fig.~\ref{fig:vx_fx-0}.

\begin{figure}[t!]
\centering
\includegraphics[width=0.9\columnwidth]{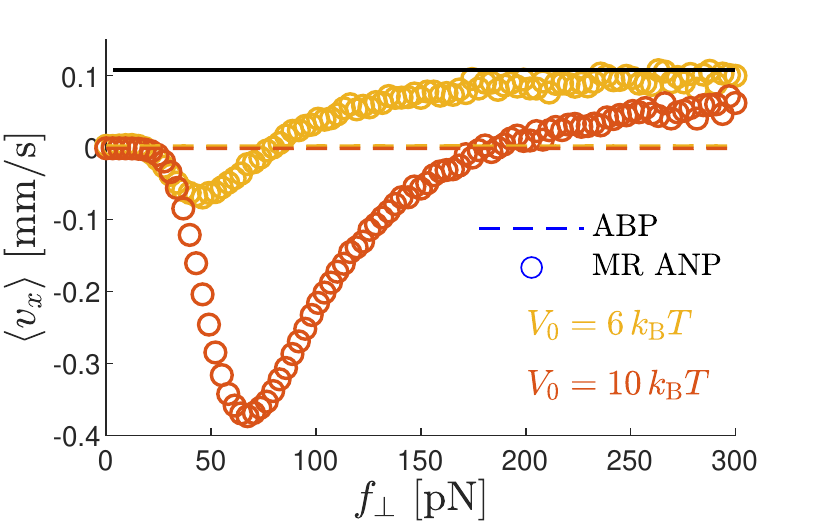} 
\caption{{\bf The ratchet effect and the velocity reversal.} The mean nanoparticle velocity $\langle v_x \rangle$ in the $x$-direction (symbols), for the case when the nanoparticle is subjected to the nonzero parallel force $\fx=0.025\;\si{pN}$. Other parameters are the same as in Fig.~\ref{fig:vx_fx-0}. The black solid line represents the velocity~\eqref{eq:vx-free} of a Brownian particle moving in a flat potential with the enhanced mobility $(\mu+\muc/2)$. The two dashed lines marking corresponding results for the ABP model overlap.}
\label{fig:vx-reversal}
\end{figure}

The velocity  $\langle v_x \rangle$ attains its maximum for moderate values of $\fy$. If $\fy\to 0$, $\langle v_x \rangle$ vanishes because a weak force does not affect the rate of the barrier-crossings. At large $\fy$, $\langle v_x \rangle$ vanishes too: When the force is extremely strong, the potential barriers of $V(x)$ have only a minor influence on the particle motion and the effect of their asymmetry becomes negligible. Such a release of nanoparticles from deep potential wells that is induced by a strong perpendicular force $\fy$ can be utilized for controlling the direction of $\langle v_x \rangle$. This is demonstrated in Fig.~\ref{fig:vx-reversal} showing a reversal of the direction of $\langle v_x \rangle$ for particles subjected to a weak force $\fx>0$ acting in the $x$-direction. 

For small and moderate $\fy$, the results shown in Fig.~\ref{fig:vx-reversal} resemble those in Fig.~\ref{fig:vx_fx-0}: MR~ABP experiences the ratchet effect and moves on average with $\langle v_x \rangle<0$. However, at larger $\fy$, as the barriers of $V(x)$ become less significant for the MR~ANP dynamics, the direction of the net particle motion is reversed, and $\langle v_x \rangle >0$. Eventually, as $\fy\to\infty$, $\langle v_x \rangle$ converges towards the solid line given by 
\begin{equation}
\label{eq:vx-free}
\langle v_x^{(0)} \rangle =\left(\mu + \frac{\muc}{2} \right) \fx ,
\end{equation}
which equals to the mean velocity of an overdamped Brownian particle diffusing in a flat potential [$V(x)=0$] with the enhanced mobility $\mueff = \mu + \muc/2$, cf.\ the mobility in Eq.~\eqref{eq:Langevin-x-approx}. 

Contrary to this complex behavior, $\langle v_x \rangle$ in the ABP model does not depend on $\fy$ and is vanishingly small because the ABP spends most of the time being trapped near the minima of $V(x)$ regardless of the value of $\fy$. 

\begin{figure}[t!]
\centering
\includegraphics[width=0.9\columnwidth]{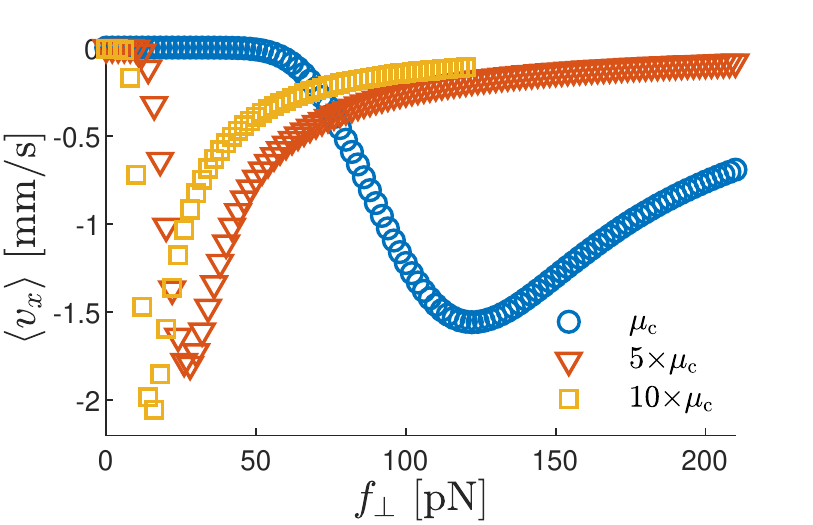} 
\caption{{\bf Impact of $\muc$ on the ratchet effect.} The mean velocity $\langle v_x \rangle$ of MR~ANP in the $x$-direction for $\fx=0$, $V_0 = 20\; \kB T$, and three values of $\muc$. Circles represent results for $\muc$ obtained from~\eqref{eq:Dc}, triangles (squares) for five (ten) times times larger $\muc$. Other parameters are the same as in Figs.~\ref{fig:vx_fx-0} and~\ref{fig:vx-reversal}.}
\label{fig:vx-muc} 
\end{figure} 
Finally, let us note that the magnitude of $\langle v_x \rangle$ and the range of $\fy$ values, where the transverse ratchet effect occurs, depend on the value of the phenomenological parameter $\muc = \Dc/\kB T$ characterizing the strength of the mechanochemical coupling. This dependence is demonstrated in Fig.~\ref{fig:vx-muc}, where blue circles correspond to $\muc$ used in all other figures and obtained from the estimate~\eqref{eq:Dc} and red triangles (yellow squares) represent simulation results for $\muc$ being 5~times (10~times) larger. Hence, for a stronger mechanochemical coupling, the ratchet effect becomes more pronounced and occurs at lower values of the transverse force $\fy$. 
\vspace{0.5cm}

\noindent {\large \bf Summary and Perspectives}

\noindent The principle of microscopic reversibility is fundamental for nonequilibrium statistical mechanics and thermodynamics. Here, we have discussed the physical origins and fundamental properties of a novel ratchet mechanism resulting from the MR of the chemically-driven self-propulsion mechanism of an active nanoparticle. The mechanism enables efficient control of the velocity of a nanoswimmer propelled by chemical reactions. It is based on an interplay of the particle translational diffusion in an asymmetric periodic potential $V(x)$ and the modification of its self-propulsion speed because of the mechanochemical coupling. The resulting ratchet effect manifests itself most strikingly by the emergent net particle motion in the easy direction of the ratchet potential, when a constant force $\fy$ is applied to the particle in a perpendicular direction. The mean velocity $\langle v_x \rangle$ of this motion can be controlled by adjusting the barrier heights $V_0$ of the periodic potential and the magnitude of $\fy$. For a given $V_0$, there exists an optimal value of $\fy$, for which the mean speed $|\langle v_x \rangle|$ attains a maximum. Furthermore, if a constant force $\fx$ is applied in the $x$-direction,  the mean velocity can be reversed by changing~$\fy$. 

In addition to numerical results, we have shown that for the model parameters roughly corresponding to catalytic enzymes, the dynamics of the nanoparticle in the $x$-direction is equivalent to that of a Brownian tilting ratchet. This mapping provides an intuitive physical interpretation of the reported ratchet effect and demonstrates that it cannot occur in active dynamics lacking the mechanochemical coupling. In the present model, this coupling arises as a consequence of the microscopic reversibility of the fluctuating chemical kinetics powering nanoparticle self-propulsion. Further exploration of such coupled thermodynamic processes at the nanoscale can inspire novel designs for micromanipulation techniques, where macromolecules are acted upon by externally applied electromagnetic or/and mechanical forces. Additionally, testing such effects in experiments may contribute toward a resolution of the long-standing question of whether active enzymes self-propel or not.  
\vspace{0.5cm}

\noindent {\large \bf Acknowledgments}

\noindent We acknowledge financial support from the Portuguese Foundation for Science and Technology (FCT) under Contracts nos.\ PTDC/FIS-MAC/5689/2020,  UIDB/00618/2020, and UIDP/00618/2020 and from the Department of Physics and Mathematics at Nottingham Trent University (grant no.\ 01/PHY/-/X1175). A.R.\ gratefully acknowledges financial support from the Czech Science Foundation (project no.\ 20-24748J). Computational resources were provided by the e-INFRA CZ project (ID:90254), supported by the Ministry of Education, Youth and Sports of the Czech Republic. 
\vspace{0.5cm}

%


\onecolumngrid
\newpage
\renewcommand{\theequation}{S\arabic{equation}}
\renewcommand{\thefigure}{S\arabic{figure}}
\setcounter{equation}{0}
\setcounter{figure}{0}

\begin{center} 
{\large\bf Supplementary Information:\texorpdfstring{\\[0.5ex]}{} Mechanochemical Active Ratchet}\\[2ex]
Artem Ryabov$^{1,2,3}$ and Mykola Tasinkevych$^{2,3,4,5}$\\[1ex]
$^1$\textit{Charles University, Faculty of Mathematics and Physics, Department of Macromolecular Physics,\\ V Hole\v{s}ovi\v{c}k\'ach 2, CZ-18000 Praha 8, Czech Republic}\\[1ex]
$^2$\textit{Departamento de F\'{\i}sica, Faculdade de Ci\^{e}ncias, Universidade de Lisboa,\\ 1749-016 Lisboa, Portugal}\\[1ex]
$^3$\textit{Centro de F\'{i}sica Te\'{o}rica e Computacional, Faculdade de Ci\^{e}ncias,  Universidade de Lisboa,\\ 1749-016 Lisboa, Portugal}\\[1ex]
$^4$\textit{SOFT Group, School of Science and Technology, Nottingham Trent University,\\ Clifton Lane, Nottingham NG11~8NS, United Kingdom}\\[1ex]
$^5$\textit{International Institute for Sustainability with Knotted Chiral Meta Matter,\\ Hiroshima University, Higashihiroshima 739-8511, Japan}
\end{center}

\setstretch{1.5} 
\noindent This Supplemental Information provides additional technical details about Brownian dynamics simulations that were used to obtain the results presented in the main text. Also, the simplified approximate Langevin equation is derived.
\vspace{1ex}

\noindent {\bf \large Supplementary Note 1: Langevin Equations in Dimensionless Units}

\noindent We consider an active nanoparticle diffusing in the force field  
\begin{equation}
\bm F(\rvec) = (F(x),\fy),  
\end{equation} 
with the constant $y$-component $\fy$ and the $x$-component being 
\begin{equation}
F(x) = \fx - \frac{\dd V}{\dd x} = \fx 
- \frac{\pi V_0}{\lambda } \left[2 \cos\left(\frac{2\pi x}{\lambda} \right) 
+ \cos\left(\frac{4\pi x}{\lambda}\right) \right] ,
\end{equation}
where $V(x)=V(x+\lambda)$ is the external $\lambda$-periodic potential. The system of Langevin equations governing dynamics of the particle's center of mass position~$\rvec(t)=(x(t),y(t))$, and the orientation $\phi(t)$ is 
\begin{subequations}
\label{eqS:Langevin}
\begin{align}
\label{eqS:Langevin-x}
& \frac{\dd x}{\dd t} = \ua(x,t) \cos\phi(t) + \mu F(x)  +  \sqrt{2D}\, \xi_x(t) ,\\    
& \frac{\dd y}{\dd t} = \ua(x,t) \sin\phi(t) + \mu \fy  +  \sqrt{2D}\, \xi_y(t) ,\\
& \frac{\dd \phi}{\dd t} =  \sqrt{2\Dr}\, \xir(t),
\end{align} 
\end{subequations}
where 
\begin{equation}
\ua(x,t) = u + \muc \left[ F(x)\cos\phi(t) + \fy \sin\phi(t) \right] + \sqrt{2\Dc}\,\xic(t),
\end{equation}
is the active speed and the Gaussian white noise processes $\xi_\alpha(t)$ satisfy 
\begin{align} 
\label{eqS:noise-mean}
\langle \xi_\alpha(t) \rangle & =0, \\ 
\label{eqS:noice-variance}
\langle \xi_\alpha(t) \xi_\beta(t') \rangle & = \delta_{\alpha\beta} \delta(t-t'),    
\end{align}
$\alpha, \beta \in \{x,y,\textrm{r},\textrm{c}\}$. 

As the natural unit of length we choose the period $\lambda$ of the external potential $V(x)$. (Alternatively, one could use the hydrodynamic radius of the nanoparticle.) As the natural unit of time we adopt the characteristic timescale of the fastest process in the model: the rotational diffusion. This characteristic timescale is given by $1/\Dr$. 
Expressing coordinates and time in these units yields   
\begin{align} 
& X = \frac{x}{\lambda} , \\ 
& Y = \frac{y}{\lambda} , \\
& \tau = \Dr t. 
\end{align} 

Let us now derive the Langevin equations for the dimensionless coordinates $X$ and $Y$ with respect to the time $\tau$. These equations will be formulated by means of dimensionless forces, dimensionless noises, and the dimensionless active speed. 

The dimensionless $x$-component of the force $\Ftr(X)$, the dimensionless constant $x$-component $\fxtr$, and the dimensionless $y$-component $\fytr$ are respectively defined as  
\begin{align}
\Ftr(X) & = \frac{\lambda F(\lambda X)}{\kB T}= -\, \pi \Votr \left[2 \cos\left( 2\pi X \right) + \cos\left(4\pi X\right) \right] + \fxtr, \\ 
\fxtr & = \frac{\lambda \fx}{\kB T},\\ 
\fytr & = \frac{\lambda \fy}{\kB T},
\end{align}
where the dimensionless amplitude of the periodic potential  
\begin{equation}
\Votr = \frac{V_0}{\kB T} 
\end{equation} 
is expressed in units of $\kB T$. 

Gaussian white noises are transformed according to 
\begin{equation}
\zeta_\alpha(\tau) = \frac{1}{\Dr}\, \xi_\alpha\!\left(\frac{\tau}{\Dr}\right),  
\end{equation}
$\alpha \in \{x,y,\textrm{r},\textrm{c}\}$. The transformed dimensionless processes $\zeta_\alpha(\tau)$  are independent, each of them has the mean value zero and they are delta-correlated in dimensionless time, similarly to the original noises $\xi_\alpha(t)$ satisfying Eqs.~\eqref{eqS:noise-mean} and~\eqref{eqS:noice-variance}. Specifically, 
$\langle \zeta_\alpha(\tau) \rangle =0$ follows from~\eqref{eqS:noise-mean}, and   
$\langle \zeta_\alpha(\tau) \zeta_\beta(\tau') \rangle = \delta_{\alpha\beta} \delta(\tau-\tau')$ is justified employing the identity $\delta(\tau/\Dr)=\Dr \delta(\tau)$.

In course of the transition to the dimensionless units, the diffusion coefficients $D$ and $\Dc$ are scaled by the factor $\lambda^2\Dr$ giving two dimensionless quantities 
\begin{align}
\label{eqS:Dtr-def}
& \Dtr = \frac{D}{\lambda^2 \Dr},  \\  
\label{eqS:Dctr-def}
& \Dctr = \frac{\Dc}{\lambda^2 \Dr} . 
\end{align}
Furthermore, the combination $\lambda\Dr$ sets the unit of velocity, yielding  
\begin{align}
\uatr(X,\tau) & = \frac{ \ua(\lambda X, \tau/\Dr)}{\lambda \Dr } = \utr + \Dctr \left[ \Ftr(X)\cos\phi(\tau) + \fytr \sin\phi(\tau) \right]  + \sqrt{2\Dctr} \,\zetac (\tau),  \\ 
\label{eqS:utr-def}
\utr &= \frac{ u}{\lambda \Dr}. 
\end{align} 

The Langevin equations~\eqref{eqS:Langevin}, when expressed using these dimensionless quantities, read 
\begin{subequations} 
\label{eqS:Langevin-transformed}
\begin{align}
& \frac{\dd X}{\dd \tau}  = \uatr(X,\tau) \cos\phi(\tau) + \Dtr \Ftr(X) + \sqrt{2\Dtr}\, \zeta_x(\tau) ,\\    
& \frac{\dd Y}{\dd \tau}  = \uatr(X,\tau) \sin\phi(\tau) + \Dtr \fytr  +  \sqrt{2\Dtr}\, \zeta_y(\tau) ,\\
& \frac{\dd \phi}{\dd \tau}  =  \sqrt{2}\, \zetar(\tau).
\end{align} 
\end{subequations}

Numerical values of the dimensionless model parameters $\Dtr$, $\Dctr$, and $\utr$, defined in Eqs.~\eqref{eqS:Dtr-def}, \eqref{eqS:Dctr-def}, and~\eqref{eqS:utr-def}, follow from Eqs.~(6)-(10) in the main text:  
\begin{align}
\label{eqS:Dtr}
\Dtr & = 0.03,  \\  
\label{eqS:Dctr}
\Dctr & \approx 0.0022, \\ 
\label{eqS:utr}
\utr & \approx 1.7 \times 10^{-6}.
\end{align}

\noindent {\bf \large Supplementary Note 2: Brownian Dynamics Simulations}

\noindent In our numerical simulations, we obtained approximate solutions of Eqs.~\eqref{eqS:Langevin-transformed} by implementing the Euler–Maruyama method \cite{Kloeden/Platen:1992}.  
Having values of $X(\tau)$, $Y(\tau)$, and $\phi(\tau)$, the method approximates the corresponding values of $X(\tau+\Delta\tau)$, $Y(\tau+\Delta\tau)$, and $\phi(\tau+\Delta\tau)$ according to the recurrence relations
\begin{subequations}
\label{eqS:Langevin-numerics} 
\begin{align} 
X(\tau+\Delta\tau) & = X(\tau) + \left[ \uatr(X(\tau),\tau)\cos\phi(\tau) + \Dtr\Ftr(X)\right] \Delta\tau + \sqrt{2\Dtr\Delta\tau}\, \Nx ,\\    
Y(\tau+\Delta\tau) & = Y(\tau) + \left[ \uatr(X(\tau),\tau)\sin\phi(\tau)+\Dtr\fytr\right] \Delta\tau +  \sqrt{2\Dtr\Delta\tau}\, \Ny ,\\ 
    \phi(\tau+\Delta\tau) & = \phi(\tau) +  \sqrt{2\Delta\tau}\, \Nr,
\end{align} 
\end{subequations}
where 
\begin{equation}
\ua(X(\tau),\tau) = \utr + \Dctr \left[ \Ftr(X(\tau))\cos\phi(\tau) + \fytr \sin\phi(\tau) \right] + \sqrt{\frac{2\Dctr}{ \Delta\tau}} \,\Nc ,
\end{equation}
and $\mathcal{N}_\alpha(0,1)$, $\alpha \in \{x,y,\textrm{r},\textrm{c}\}$, are independent and identically distributed random numbers drawn from the normal distribution with zero mean and unit variance. 

In all simulations, at the initial time $\tau=0$, we set $X(0)=Y(0)=0$. The initial particle orientation is assumed to be random, i.e.,  $\phi(0)$ is a random variable homogeneously distributed within the interval $[0,2\pi)$.

Since $\Dtr$ and $\Dctr$, and $\utr$ are much smaller than one, see Eqs.~\eqref{eqS:Dtr}-\eqref{eqS:utr}, the translational diffusion and the active motion take place on significantly slower timescales than the rotational diffusion does. Therefore, to achieve a satisfactory precision of the numerical scheme~\eqref{eqS:Langevin-numerics}, it is necessary to choose $\Delta \tau$ such that the rotational diffusion process is simulated precisely enough. The difference in time scales then would ensure that the particle displacements caused by the two slow processes are approximated well too. 

We have found that $\Delta\tau=10^{-3}$, used to generate all simulation results displayed in the main text, is small enough for this purpose. To verify that the results of simulations do not depend on $\Delta\tau$, we performed simulations for a few selected parameter sets and $\Delta\tau=10^{-4}$ and $\Delta\tau=10^{-5}$. 

As a consequence of the timescale separation in our model, long overall simulation time $\tau_{\rm max}$ (compared to $\Delta \tau$) must be chosen to generate a noticeable translational motion of the particle across several wells of the potential $V(x)$. The long $\tau_{\rm max}$ is also necessary to eliminate all transient effects. We have used $\tau_{\rm max}=10^3$. 

For each given set of model parameters, the mean particle velocity $\langle v_x \rangle$ in the $x$-direction has been obtained as the long-time average of the particle displacement along the $x$-axis: 
\begin{equation}
\label{eqS:vx}
\langle v_x \rangle = \frac{1}{N_{\rm tr}}\sum_{i=1}^{N_{\rm tr}} \frac{x_i(\tau_{\rm max})}{\tau_{\rm max}} ,
\end{equation}
where $x_i(\tau_{\rm max})$ corresponds to the final point of the $i$-th trajectory and $N_{\rm tr}$ is the total number of trajectories. For all reported data points, we used $N_{\rm tr}=10^4$. We have checked that increasing the value of $N_{\rm tr}$ to $10^5$ and $10^6$ (and of $\tau_{\rm max}$ to  $\tau_{\rm max}=10^4$ and $10^5$) does not affect the results. These stability checks were performed for larger values of $\Votr$ and $\fytr$ where larger numerical errors are expected to occur. 
\vspace{0.5cm}

\noindent {\bf \large Supplementary Note 3: Approximate Langevin Equation for the Longitudinal Coordinate}

\noindent The Langevin equation~\eqref{eqS:Langevin-x} for the longitudinal coordinate $x$ is given by 
\begin{equation} 
\label{eqS:Langevin-x-explicit}
\frac{\dd x}{\dd t} = \left\{ u + \muc \left[ F(x)\cos\phi(t) + \fy \sin\phi(t) \right] + \sqrt{2\Dc}\,\xic(t) \right\} \cos\phi(t) + \mu F(x)  +  \sqrt{2D}\, \xi_x(t) ,
\end{equation}
which we can rearrange to a compact form 
\begin{equation}
\label{eqS:Langevin-x-explicit-rearranged}
\frac{\dd x}{\dd t} =  u \cos\phi(t) + \frac{ \muc \fy}{2} \sin[2\phi(t)] + 
\left[\mu + \muc\cos^2\phi(t) \right] F(x) 
 +  \sqrt{2\kB T \left[ \mu + \muc\cos^2\phi(t)  \right]}\, \xi(t) ,
\end{equation} 
where $\xi(t)$ is the Gaussian white noise process satisfying  
$\langle \xi(t)\rangle = 0$ and $\langle \xi(t)\xi(t')\rangle = \delta(t-t')$. 
Equation \eqref{eqS:Langevin-x-explicit-rearranged} is equivalent to Eq.~\eqref{eqS:Langevin-x-explicit}. In deriving Eq.~\eqref{eqS:Langevin-x-explicit-rearranged} from Eq.~\eqref{eqS:Langevin-x-explicit}, we have employed the identity $\sin( 2\phi) = 2 \sin \phi \cos \phi$, the fluctuation-dissipation theorems $D=\mu \kB T$, $\Dc=\muc \kB T$ from the main text, and the property of Gaussian random variables \cite{Lemons:2002}: 
\begin{equation}
\sqrt{\sigma^2_1}\;\mathcal{N}_1(0,1) + \sqrt{\sigma^2_2}\; \mathcal{N}_2(0,1) =  
\sqrt{\sigma^2_1+\sigma^2_2}\; \mathcal{N}_3(0,1) ,
\end{equation}
where $\mathcal{N}_\alpha(0,1)$, $\alpha \in \{1,2,3\}$, are independent and identically distributed normal random variables with zero mean and unit variance. 

For the values of parameters $u$, $\mu$, $\muc$, $\fy$, and $\Dr$ used in the main text, Eq.~\eqref{eqS:Langevin-x-explicit-rearranged} can be considerably simplified. This is done in two steps. 

First, for $u \lesssim 1000\;\si{nm/s}$
we can neglect the term proportional to $u$ in Eq.~\eqref{eqS:Langevin-x-explicit-rearranged}. This approximation is valid for active particles with smaller hydrodynamic radii $\RH$. It results from the large value of $\Dr$, $\Dr \sim 1/\RH^3$, Eq.~(7) in the main text. Thus, for $u\approx 10\;\si{nm/s}$, used in all figures in the main text, Eq.~\eqref{eqS:Langevin-x-explicit-rearranged} reduces to 
\begin{equation}
\frac{\dd x}{\dd t} \approx  \frac{ \muc \fy}{2} \sin[2\phi(t)] + \left[\mu + \muc\cos^2\phi(t) \right] F(x) 
 +  \sqrt{2\kB T \left[ \mu + \muc\cos^2\phi(t)  \right]}\, \xi(t). 
\end{equation} 

Second, according to Eqs.~(6) and (8), $\Dc /D = \muc/\mu \approx 0.073$, i.e.,  in the expression
$\left[ \mu + \muc\cos^2\phi(t)  \right]$, the term $\muc\cos^2\phi(t)$ is negligible as compared to $\mu$. 
Numerically we have found, that the results obtained after the approximation 
\begin{equation}
\label{eqS:muc-approximation}
\mu + \muc\cos^2\phi(t)     
\approx \mu + \frac{\muc}{2},
\end{equation} 
i.e., when we approximate the nonnegative function by its average, $\cos^2\phi(t) \approx 1/2$, fit the simulated data slightly better than the curves obtained by the simple approximation $\left[ \mu + \muc\cos^2\phi(t)  \right] \approx \mu$, the difference being of the order of percents. Employing the approximation~\eqref{eqS:muc-approximation}, gives us
\begin{equation} 
\label{eqS:Langevin-approximate}
\frac{\dd x}{\dd t} \approx  \frac{ \muc \fy}{2} \sin[2\phi(t)] + 
\left(\mu + \frac{\muc}{2} \right) F(x) 
 +  \sqrt{2\kB T \left( \mu + \frac{\muc}{2} \right)}\, \xi(t) .
\end{equation}
Setting further $\fx=0$ yields Eq.~(11) in the main text. 

\begin{figure}[ht!]
\centering
\includegraphics[width=0.9\columnwidth]{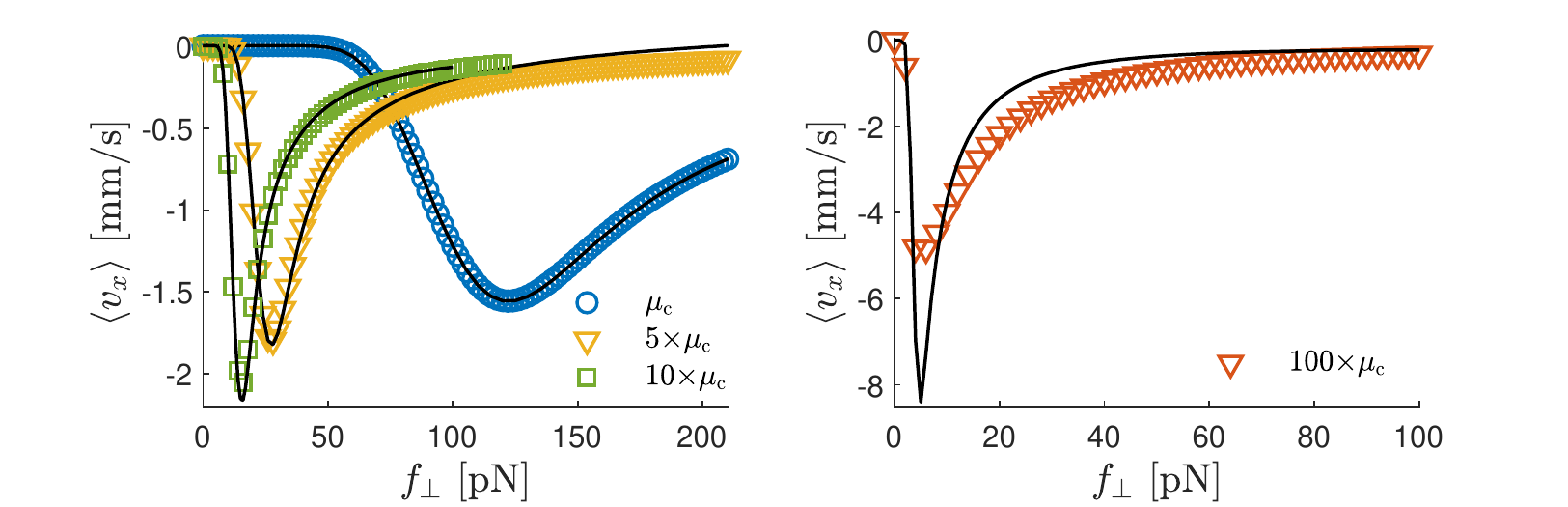} 
\caption{Mean velocity $\langle v_x \rangle$ in the $x$-direction, Eq.~\eqref{eqS:vx}, for $\fx=0$, $V_0 = 20\; \kB T$, and four values of the parameter $\muc$. Other model parameters are the same as in Figs.~2 and~3 in the main text. Symbols (lines) represent results based on the numerical integration of the exact Eq.~\eqref{eqS:Langevin-x-explicit} [approximate Eq.~\eqref{eqS:Langevin-approximate}]. Left panel: circles mark results for the reference value of $\muc$ obtained from Eq.~(8) in the main text, triangles (squares) correspond to values of $\muc$ five (ten) times higher than the reference value. Right panel: $\muc$ hundred times higher than the reference value.}
\label{figS:vx-exact-vs-simplified} 
\end{figure} 

In the main text, the approximate Eq.~\eqref{eqS:Langevin-approximate} is used in the qualitative discussion of properties of the active ratchet effect. Even though no reported simulation results are based on the solution of this equation, let us now test its validity by comparing the mean velocities $\langle v_x \rangle$, Eq.~\eqref{eqS:vx}, obtained by integrating the exact Eq.~\eqref{eqS:Langevin-x-explicit} and the approximate  Eq.~\eqref{eqS:Langevin-approximate}. Such a comparison is presented in  Fig.~\ref{figS:vx-exact-vs-simplified} showing $\langle v_x \rangle$ as a function of $\fy$ for four $\muc$. In the left panel, we have used the reference value of $\muc$ obtained from Eq.~(8) in the main text and increased this reference $\muc$ by 5, 10, and 100 times. The right panel shows the extreme case of 100 times reference $\muc$. Symbols (lines) represent results based on Eq.~\eqref{eqS:Langevin-x-explicit} [Eq.~\eqref{eqS:Langevin-approximate}]. In the left panel, Eq.~\eqref{eqS:Langevin-approximate} approximates  $\langle v_x \rangle$ reasonably well. In the right panel, significant quantitative deviations between the mean velocities in the two models occur. Despite this quantitative disagreement, the overall shape of the velocity-force relations is similar. Also, both curves have a minimum at comparable values of perpendicular forces. 
\vspace{0.5cm}

\noindent {\bf \large Supplementary References}
%

\end{document}